\documentclass[12pt]{article}

\usepackage{epsfig}
\usepackage{cite}
\usepackage{graphicx}
\usepackage{amssymb}
\usepackage{amsmath}
\usepackage{latexsym}
\usepackage{feynmp}

\newcommand{\be}{\begin{eqnarray}}
\newcommand{\ee}{\end{eqnarray}}

\newcommand{\en}{\end{eqnarray}}

\newcommand{\bea}[1]{\left(\begin{array}{#1}}
\newcommand{\ena}{\end{array}\right)}

\def\norm@note#1#2{\special{}
  \ifinner{\ifdim\baselineskip=\z@
    \baselineskip18\p@\fi
    \ifhmode
    \raisebox{.5\baselineskip}[\z@][\z@]{%
      \rlap{\sf\scriptsize #2}}%
    \else\vskip-\baselineskip%
    \raisebox{-.6\baselineskip}[\z@][\z@]{%
      \rlap{\sf\scriptsize #2}}%
    \fi}%
  \else\marginpar{\raggedright\if@twoside\ifodd\c@page%
    \raggedleft\fi\fi\sf\scriptsize #1#2}%
  \fi}%

\begin{document}

\begin{titlepage}

\vskip.5cm
\begin{center}
{\huge \bf Two Higgs Models for Large $\tan\beta$ and Heavy Second
Higgs} \vskip.1cm
\end{center}
\vskip0.2cm

\begin{center}
{\bf   Lisa Randall}
\end{center}
\vskip 8pt
\begin{center}
{\it Jefferson Physical Laboratory \\
Harvard University, Cambridge, MA 02138, USA} \\
\vspace*{0.3cm} {\tt   randall@physics.harvard.edu}
\end{center}

\vglue 0.3truecm

\begin{abstract}
\vskip 3pt \noindent
    We study two Higgs models  for large $\tan\beta$ and relatively large second Higgs mass. In this limit the second heavy Higgs
    should have small vev and therefore
    couples only weakly to two gauge bosons. Furthermore, the
    couplings to down type quarks can be significantly modified (so
    long as the second Higgs is not overly heavy). Both these
    facts have significant implications for search strategies at
    the LHC and ILC. We show how an effective theory and explicit fundamental two Higgs model  approach are related
    and consider the additional constraints in the presence of  supersymmetry or $Z_2$ flavor symmetries.  We argue that
      the best tests of the two Higgs doublet potential are likely to be measurements of the light Higgs branching
     fractions. We
     show how higher dimension operators that have recently been suggested to raise the light Higgs
     mass are probably best measured and distinguished in this
     way.
\end{abstract}

\end{titlepage}

\newpage
\section{Introduction}
Two Higgs models are perhaps the simplest alternative to the
Standard Model. They are particularly important because they are
essential to low-energy supersymmetry but they of course can occur
in other models that allow a broader parameter range. The
phenomenology of the neutral Higgs sector is slightly subtle since
the angles from mass mixing are not in general the same as the
angle associated with the relative vevs. However we will see that
they generally align to a large extent when one Higgs is somewhat
heavier,
 greatly simplifying the analysis of the implications.

In this paper we explore the phenomenology of two Higgs doublet
models for large $\tan\beta$ when the lighter Higgs $h$ is light
enough so that its decays are dominantly into $b$s whereas the
second Higgs is somewhat heavy. We are motivated in part by the
analysis of \cite{rattazzi}, which performed an operator analysis
in the strongly interacting Higgs sector case to elucidate
interesting effects that can occur when the light Higgs is part of
a larger Higgs sector.

Similar considerations apply to two Higgs models, since the light
Higgs is not exactly the eaten Goldstone boson in this case
either.\footnote {Note that we use the conventional notation where
the scalar particles in the Higgs sector are called Higgs
particles. Purists might restrict this term for the linear
combination with a vacuum expectation value but since the fields
mix it is easier to call them all by this term and to distinguish
the light, heavy, and charged Higgses.} For example, we find
growth in $WW$ scattering  with energy  though it corresponds to a
higher order operator than in the strongly interacting Higgs
models considered in Ref. \cite{rattazzi} and is not the most
significant deviation from Standard Model predictions.

The modification  of the light Higgs  coupling to bottom type
quarks and charged leptons  can be significant however for two
Higgs doublet models. Although studying  deviations in the light
Higgs couplings from their Standard Model might not seem to be the
best way to study a perturbative theory where the additional
states are more likely to be kinematically accessible, we show
that for a large parameter range the heavy Higgs will probably
elude detection and precise measurements of light fields will be
the best way to test the Higgs sector.

 An
operator analysis for two Higgses  from a purely effective theory
viewpoint  was in fact completed in a recent paper \cite{wise}
where it was shown that for large Yukawa coupling of the heavy
Higgs to the down sector and small Yukawa of the heavy Higgs to
the up sector one could find significant deviations in the Higgs
partial widths for the light Higgs particle, even when the second
heavy Higgs will elude direct detection. In this paper we relate
the more conventional two Higgs analyses of Gunion and Haber
\cite{gunionhaber} to the effective theory analysis of Mantry,
Trott, and Wise \cite{wise}. We show that the conclusions reached
in that paper (namely large corrections to $b$ Yukawas and
difficulties of finding a second Higgs) apply quite generally for
large $\tan\beta$. We show that the assumption made there is in
some sense less arbitrary than it might seem in that these
characterizations apply to the Yukawa couplings for the heavy
Higgs in the large $\tan\beta$ limit.

We show however that if the dimension-4 operators respect the
$Z_2$ symmetry that guarantees a GIM mechanism that the Yukawa
modifications  are expected to be smaller since they are no longer
enhanced by $\tan\beta$. However, we will see that the effects can
still be quite significant.  We also consider the relationship
between deviations in bottom and tau branching ratios in the
various doublet Higgs models that preserve a $Z_2$ symmetry.

Finally we are motivated by recent data that point to high
supersymmetry breaking or a new physics scale motivating
considering a relatively heavy second Higgs and higher effective
dimension operators in the Higgs sector. We find that the higher
effective dimension operators of \cite{seiberg} can generate large
deviations in Higgs branching ratios and that these deviations in
Yukawas are most likely the best way to test for the new operators
they suggest. Furthermore, these deviations in Yukawas could
distinguish among the different possible higher dimension
operators that could in principle correct the light Higgs mass.

\section{General Two Higgs Analysis}
We will consider a two Higgs theory in the decoupling limit where
one of the Higgs is assumed to be light (in the regime in which
decays to $b$s dominate) and the other Higgs is assumed to be
relatively heavy. We will use two parameterizations below, and use
both $m_H$ and $M$ to  refer to the mass of the heavy Higgs.

Let us first parameterize the two Higgs Lagrangian with the
notation of Gunion and Haber \cite{gunionhaber} (see also
\cite{hhguide, chgeneral}) but using the notation $H_1$ and $H_2$
for the two Higgs bosons. We have the gauge invariant scalar
potential
\begin{eqnarray}
V&=&m_{11}^2 H_1^\dagger H_1 +m_{22}^2 H_2^\dagger H_2-[m_{12}^2
H_1^\dagger H_2+{\rm h.c.}] +{1 \over 2} \lambda_1 (H_1^\dagger
H_1)^2 + {1 \over 2} \lambda_2 (H_2^\dagger H_2)^2 \nonumber \\
&+&\lambda_3 (H_1^\dagger H_1) (H_2^\dagger H_2) + \lambda_4
(H_1^\dagger H_2) (H_2^\dagger H_1)+({1\over 2}
\lambda_5(H_1^\dagger H_2)^2 \nonumber \\ &+&(\lambda_6
(H_1^\dagger H_1)+\lambda_7(H_2^\dagger H_2))H_1^\dagger H_2 +{\rm
h. c.})
\end{eqnarray}

We take all parameters to be real and CP-conserving for
simplicity. In a supersymmetric model, these parameters take the
values
\begin{equation}
\lambda_1=\lambda_2=-\lambda_{345}={1 \over 4} (g^2+g\prime^2), \
\ \lambda_4=-{1 \over 2} g^2, \\\lambda_5=\lambda_6=\lambda_7=0
\end{equation}
Where $\lambda_{345}=\lambda_3+\lambda_4+\lambda_5$. Notice that
the last two parameters are zero in any model that respects a
$Z_2$ symmetry in the dimension-4 operators. We might expect this
to be approximately the case in any of the standard two Higgs
scenarios where an approximate $Z_2$ guarantees a GIM mechanism.
However, breaking of the $Z_2$ in the dimension-4 operators above
can exist while still not introducing overly-large flavor changing
effects \cite{wise,seiberg}. As we will see such cases will lead
to particularly interesting deviations from the Standard Model.

We now assume an appoximate $Z_2$ symmetry and follow the standard
notation and define the ratio of vevs of the two fields as
$\tan\beta=\langle H_2 \rangle /\langle H_1\rangle$, where $H_2$
is the field coupling to the top quarks and $H_1$ is the field
coupling to the bottom quarks and charged leptons (here we are
assuming a Type II model where this is the case but we will
explore later other assumptions). The angle $\alpha$ determines
the mixing angles of the Higgs fields mass eigenstates, so we have

\begin{equation}
H_1={1 \over \sqrt{2}} (\cos\alpha H-\sin\alpha L)
\end{equation}
\begin{equation}
H_2={1 \over \sqrt{2}} (\sin\alpha H+\cos\alpha L)
\end{equation}
where $H$ is the heavy Higgs field and $L$ is the light Higgs
field and we are only considering the real parts of $H_1$ and
$H_2$. Notice that with this parameterization the  fields $H$ and
$L$ have nonzero vevs, but this can be subtracted off as in
\cite{gunionhaber}.

The vevs for the two (real) Higgs fields  (neglecting higher order
terms in $(v/m_H)^2$ are given by
\begin{equation}
\langle L \rangle=v \sin (\beta-\alpha)
\end{equation}
\begin{equation}
\label{ghvev}
 \langle H \rangle= v \cos (\beta-\alpha)
\end{equation}

where
\begin{equation}
\label{cosbma}
 \cos (\beta-\alpha)\sim{\hat{\lambda} v^2 \over
m_H^2}
\end{equation}

and
\begin{equation}
\label{lambdahat}
 \hat{\lambda}={1 \over 2} \sin{2 \beta} (
\lambda_1 \cos^2\beta-\lambda_2 \sin^2\beta- \lambda_{345} \cos {2
\beta})-\lambda_6 \cos\beta \cos {3 \beta}-\lambda_7 \sin\beta
\sin {3 \beta}
\end{equation}
In the large $\tan\beta$ limit, this reduces to
\begin{equation}
\hat{\lambda}=\cos\beta(-\lambda_2+\lambda_{345})+\lambda_7
\end{equation}
and in a supersymmetric theory we would have
\begin{equation}
\hat{\lambda}=-{\cos\beta \over 2} (g^2+g\prime^2)\sim-0.3
\cos\beta
\end{equation}

Alternatively in the limit that $\tan \beta$ is large one can just
solve for $\sin\alpha$ (using the mass matrices from
\cite{gunionhaber}) (see Eq. (\ref{massmatrix}) below) to find
\begin{equation}
\label{cos} \sin\alpha \sim -\cos\beta+\lambda_7 (v^2/M^2).
\end{equation}
Expanding out $\cos(\beta-\alpha)$ in Eq. (\ref{cosbma}) gives
this same expression.

The equations above  show that the masses and vevs are aligned up
to $O(v^2/M^2)$ corrections. The heavy Higgs  gets a vev through
its interaction with the light Higgs field that has acquired a
larger vev. This would have been more manifest with different
notation. For example, the vev of $H_1$ is proportional to
$v\cos\beta$ whereas the coefficient of $L$ is $-\sin\alpha$, so
it might have been natural when the quartic term doesn't dominate
the mixing to have started with the rotated angle  $-\sin\alpha
\to \cos\alpha$ in the first place.

Notice that the equations above imply that $\alpha\sim
\beta-\pi/2$ in the extreme decoupling limit where $\tan\beta
\lambda v^2 /m_H^2<1$. Although perhaps not as likely to be
physically relevant, we also consider  the opposite limit, where
the off-diagonal term in the mass matrix changes sign. In this
case, the above results still hold for $\alpha$ and the vev of the
heavy field, although when the $\lambda_7$ term dominates
$\sin\alpha$ reverses sign.

The answer above suffices over the entire parameter range but for
completeness we compare the result to that of \cite{gunionhaber}
in this limit, noting  that the intermediate results can depend on
convention.\footnote{We thank Howie Haber for discussions on this
limit.} Ref. \cite{gunionhaber} gave $\alpha\sim\pi/2-\beta$
\cite{gunionhaber}, $\sin\alpha\sim\cos\beta$, which is the result
when the initial conventions for the Lagrangian do not account for
$\cos\beta>0$. Minimizing the potential with respect to $\phi_1$
(substituting in the assumed form for the $H_1$ and $H_2$ vevs)
yields the equation \cite{gunionhaber}
\begin{equation}
\label{min}
 m_{11}^2=m_{12}^2\tan\beta-{1 \over 2} v^2
\left(\lambda_1 \cos^2\beta+\lambda_{345} \sin^2\beta +3 \lambda_6
\sin\beta\cos\beta+\lambda_7 \sin^2 \beta \tan\beta \right)
\end{equation}
When the $\lambda_7$-dependent-term dominates, one needs negative
$\cos\beta$ to satisfy this equation. However, according to the
\cite{gunionhaber} convention $\beta$ is always between $0$ and
$\pi/2$.  In order to maintain $\cos\beta>0$ and $m_{11}^2>0$
(when $\lambda_7>0$), we need to change the sign of $H_1$.  With
this sign change,  we can directly solve for $\sin\alpha$ (in the
large $\tan\beta$ limit to find $\sin\alpha\sim
\cos\beta-\lambda_7 (v/M)^2$. In this case we can evaluate
$\cos(\beta-\alpha)$ to find approximately $2\cos\beta$ as above,
but the more useful quantity would be the quantity that appears in
the $H$ vev. Because we have changed the sign of $H_1$, we see
that the vev of $H$ is related to $-\cos(\alpha +\beta)$, and this
again evaluates to $\lambda_7 (v/M)^2$. Alternatively had we taken
a convention where we also changed the sign of $\sin\alpha$, we
would have obtained the answers we did for small $\lambda_7$
above.  In either way of proceeding, the vev of the heavy field
and $\sin\alpha$ (up to an unphysical sign) take the same form,
even when $\lambda_7 \tan \beta (v/M)^2>1$. These are the
physically relevant quantity that enter the heavy Higgs coupling
to two gauge bosons and the Yukawas. So our results
(\ref{ghvev}),(\ref{cos}) for the vev and mixing angle apply over
the entire parameter range.

An alternative approach to a two Higgs model with a heavy Higgs is
to take an effective theory approach as considered in \cite{wise}.
Ref. \cite{wise} does not assume the existence of  a $Z_2$
symmetry (in fact $H_1$ and $H_2$ are never mentioned) so the
Yukawa couplings of the heavy Higgs can be taken as free
parameters, but the parameters were chosen to be consistent with
small FCNC (that is minimal flavor violation \cite{mfv}, assuming
only a single Yukawa matrix structure for up type quarks and
another for down type quarks). For simplicity in comparing to
their results we will call their light Higgs $H$ and their heavy
Higgs $S$ as in Ref. \cite{wise} (but note that $H$ is now the
{\it light} field and $S$ is a doublet). Their Lagrangian is:

\begin{eqnarray}
V(H,S)&=&{\lambda \over 4} (H^\dagger H-{v^2 \over 2})^2+M^2
S^\dagger S +{\lambda_S \over 4} (S^\dagger S)^2+[g_1(S^\dagger
H)(H^\dagger H)+{\rm h. c.}]\nonumber \\ &+& g_2(S^\dagger S)(H
^\dagger H)  +\left[g_{2a}(S^\dagger H)(S^\dagger S)+{\rm
h.c.}\right]\nonumber \\ &+& g_{2b}(S^\dagger H)(H^\dagger
S)+[g_3(S^\dagger S)(S^\dagger H)+{\rm h.c.}].
\end{eqnarray}

Note that $g_1$ and $g_2$ are couplings completely independent of
gauge couplings; we have kept the notation  of \cite{wise} for
simplicity. Secondly, notice that all the same types of terms
appear in the non-effective theory $H_1$ $H_2$ Lagrangian aside
from the quadratic mass mixing term. However, since $H_2$ and $H$
(from \cite{wise}) are not identical, the $g$s would be a function
of various couplings in the Lagrangian above. We can expand in
terms of $\cos\beta$ to solve for one field in terms of the other.

To simply relate couplings we can consider the small $\cos\beta$
limit. In this limit, the heavy Higgs is approximately $H_1$ and
the light Higgs is approximately $H_2$. In this limit we can
expand to see that $g_1=\hat{\lambda}$.  For the more exact
result, we can expand
 $H_1$ and $H_2$ in terms of $H$ and
$L$ and include the additional $Z_2$-violating
$\cos\beta$-suppressed terms to find $g_1=\hat{\lambda}$. For
simplicity, we concentrate on the $\lambda_7$ term below.

More relevantly for physical consequences, we can  relate the vevs
and in particular the heavy Higgs vev in the two pictures. Ref.
\cite{wise} had  \footnote{Note the sign correction to
\cite{wise}.}
\begin{equation}
\label{svev}
 \langle {S \sqrt{2}} \rangle = -{g_1 v^3 \over
2M^2}
\end{equation}
(again we are working to leading order in $(v/M)^2$).
 Notice that $H$ and $L$ are
real fields in the first analysis so that the relevant field to
compare to is $\sqrt{2}S$ (ignoring the other Higgs components,
where $H$ is the heavy Higgs in the fundamental theory).

Recall that when $\cos \beta$ is small, $\lambda_7 \sim g_1$. We
see that the two values of the expectation value, though having
the same parametric dependence, differ by a factor of -2. The
reason for this is that the fundamental Higgs analysis uses the
mass eigenstates for the full mass matrix, whereas the effective
theory analysis did not use mass eigenstates once the
$g_1$-dependent quartic term is included. The physical mass
eigenstate is $S+3g_1/2 (v/M)^2H$ and has vev that agrees with the
vev for the fundamental theory when $g_1=\hat{\lambda}$.  Eq.
(\ref{min}) tells us  that without the $m_{12}^2$ term that
$\cos\beta$ would agree with Eq. (\ref{svev}) above. However, the
$g_1$ quartic (or in Ref. \cite{gunionhaber} the
$\lambda_7$-dependent quartic) also contributes to mass mixing, so
the the heavy {\it physical} mass eigenstate   has the vev cited
in Eq. (\ref{ghvev}).

To further understand this result, it is of interest to consider
the contributions of the quadratic and quartic terms to both mass
mixing and vev. Had the only mixing term between $H_1$ and $H_2$
been a mass term, one could in fact simultaneously diagonalize the
mass and vev. However, the relative mass squared coming from the
quartic is $3/2 g_1 (v/M)^2$, whereas the vev contribution to the
linear term is $g_1/2 (v/M)^2 v$. So a piece of the quartic can be
absorbed into $M_A^2$ as is done in \cite{gunionhaber}. That is,
the mass matrix takes the form

\begin{equation}
\label{massmatrix}
 M^2= M_A^2 \left(
\begin{tabular}{r r }

$\sin^2\beta$&$-\sin\beta\cos\beta$\\
$-\sin\beta\cos\beta$&$\cos^2\beta$\\
\end{tabular} \right)+{\cal B^{\rm 2}}\
\end{equation}
where $M_A^2={m_{12}^2 \over \sin\beta\cos\beta} -{1 \over 2} v^2
(2\lambda_5+\lambda_6 \tan\beta^{-1} +\lambda_7 \tan\beta)$ and
the off-diagonal part of ${\cal B^{\rm 2}}$ contains a term
$\lambda_7 v^2 \sin^2\beta$.  After full diagonalization, one is
left with   the vev of the heavy Higgs eignenstate
$\cos(\beta-\alpha)=\hat{\lambda} (v/M)^2$ as we found above.



Before closing this section, we remark on how small the VEV of the
second field is likely to be. This makes the coupling to two $W$s
very suppressed, which is essentially why the heavy Higgs search
is quite difficult as we will discuss further shortly. In the next
section we discuss the deviation of the {\it light} Higgs Yukawa
from its Standard Model value. For a large range of parameters,
this is the likely to be the best way to search for evidence of a
second Higgs.

\section{Yukawas}

Given the expressions for $H_1$ and $H_2$ in terms of $H$ and $L$,
we can work out the Yukawas for the light and heavy fields to the
up and down type quarks. In this section we will focus on
precision light Higgs measurements and study the deviation of
Higgs couplings to fermions from their Standard Model values. We
will first consider Type II models (as in the MSSM) in which one
Higgs gives mass to charged leptons and down-type quarks and the
other Higgs gives mass to up-type quarks. We then have (in
relation to the standard Yukawa couplings) \cite{gunionhaber,
pandita}

\begin{equation}
\label{hy1}
 h D \bar{D}: -{\sin\alpha \over \cos \beta}=\sin
(\beta-\alpha)-\tan\beta \cos (\beta - \alpha)
\end{equation}

\begin{equation}
h U \bar{U}: {\cos \alpha \over \sin \beta}
=\sin(\beta-\alpha)+\cot\beta\cos(\beta-\alpha)
\end{equation}

Note that both of these are of order unity when the second Higgs
is heavy and $\cos(\beta-\alpha)$ is small, as they should be in
the decoupling limit. We also see that the corrections term in the
down-type Yukawa can grow with $\tan\beta$ and be quite large.

Ref. \cite{wise} did not assume a $Z_2$ symmetry but did assume
minimal flavor violation. Note that this is more general in that
with $Z_2$ symmetry, there are only three distinct possibilities,
in which either the same Higgs or orthogonal Higgses couple to up
and down type quarks respectively. With only MFV, one can in
principle define the Higgs that couples to up-type quarks and the
one coupling to down-type quarks as $H_2$ and $H_1$, but these are
not necessarily either the same or orthogonal so there is a
continuum of possibilities. However, we will see that only the
down-type Yukawa deviations are likely to be significant when
$\tan\beta$ is large so the difference isn't necessarily
significant.

 The authors of Ref. \cite{wise}  defined
parameters $\eta_d$ and $\eta_u$ which when multiplied by the
light Higgs Yukawas of the effective theory gave the heavy Higgs
Yukawas.  In  terms of the quark masses (and including both the
light and heavy Higgs vev contributions), the Yukawa couplings of
the heavy Higgses are therefore
\begin{equation}
-\eta_d \sqrt{2} \bar{d} {m_d \over 1+\sqrt{2} \eta_d{\langle S
\rangle \over v}} d \,
\end{equation}
and similarly for up quarks, where this expression includes the
$S$ vev contribution to the quark masses.

Ref. \cite{wise} considered the possibility that $\eta_d$
is large and $\eta_u$
is small. By integrating out the heavy Higgs (and including its
vacuum expectation value contribution to the quark masses), they
found a light Higgs Yukawa coupling \footnote{Notice a sign
correction from Ref. \cite{wise}. This sign has physical
consequences since it is the deviation from the Standard Model
value.}
\begin{equation}
\label{wiseyukawa}
 { 1-{3 \over 2} g_1 \eta_d ({v \over M})^2   \over 1-{1 \over 2}
 g_1 \eta_d ({v \over M})^2}
\end{equation}
They noted that the correction is large when $\eta_d$ is big,
which is clearly similar to the observation we made above for
large $\tan\beta$.

 We now show  the similarity of these large Yukawa corrections is not a coincidence and
 that such a scheme is a generic
prediction of large $\tan\beta$.\footnote{Of course the Lagrangian
in \cite{wise} is more general, and $\tan\beta$ is not even
defined in the absence of a $Z_2$ symmetry \cite{davidson}. Our
point is that the particularly interesting case of large
$\tan\beta$ is an example of this type of parameter regime.}  This
large deviation has significant implications for the search for a
second Higgs.

The heavy Higgs coupling to down type quarks (again in relation to standard Yukawas) is given by
\begin{equation}
\label{16}
 {\cos \alpha \over \cos
\beta}=\cos(\beta-\alpha)+\tan\beta\sin(\beta-\alpha)\approx
\tan\beta
\end{equation}

whereas the coupling to up quarks is given by
\begin{equation}
{\sin\alpha \over
\sin\beta}=\cos(\beta-\alpha)-\cot\beta\sin(\beta-\alpha)<<1
\end{equation}

So we see that large $\tan\beta$ naturally yields a large Yukawa
coupling of the heavy Higgs to down quarks and a suppressed
coupling to up type quarks. We can see this directly in equation
(\ref{16}) noting that $\cos\alpha$ is very close to $\sin\beta$
(which follows from $\cos (\beta -\alpha)$ being small), so that
the value of $\eta_d$ that this model matches onto is very close
to $\tan\beta$. This follows from the original $Z_2$ symmetry,
which favors the heavy Higgs which is approximately $H_1$ coupling
to down quarks and the light Higgs which is approximately $H_2$
coupling to light quarks.


 For completeness and to elucidate the origin of this correction
 we
 do the matching for the heavy Higgs down-type Yukawa
couplings more exactly in order to compare the two formulations.
Again  when comparing the results we need to take into account
that the \cite{wise} analysis based on the effective theory
doesn't use the fully diagonalized states. So the Yukawa for the
light not quite diagonalized field in the fundamental theory would
be approximately $-\sin\alpha+{3 \over 2} \lambda_7
(v/M)^2\sim\cos\beta+{1 \over 2} \lambda_7 (v/M)^2$. So we
identify
\begin{equation}
\eta_d={\tan\beta \over 1+{\lambda_7 \over 2} \tan \beta \left({v
\over M}\right)^2}
\end{equation}
from which we conclude
\begin{equation}
{1-{3 \over 2} g_1\eta_d (v/M)^2 \over 1-{1\over 2}g_1\eta_d
(v/M)^2}=1-\lambda_7 \tan\beta(v/M)^2
\end{equation}
(where we have made the approximate identification $\lambda_7$
with $g_1$) which agrees with Eq. (\ref{hy1}). Notice that when
integrating out $S$ to determine the Yukawa, one is effectively
accounting for the mass mixing so in this case the results in the
two formulations agree. That is, the Yukawa in Eq.
(\ref{wiseyukawa}) is really the Yukawa for the physical light
Higgs. Also note that the different $Z_2$-violating quartic
contribution to the mixing and the vev leads to the  correction to
the light Higgs Yukawa.

 We see in either formulation that the correction
 can be quite
large in the large $\tan\beta$ (or large $\eta_d$)
regime.
For large $\tan \beta$
and not overly heavy Higgs mass, we can have large corrections to
the bottom and tau (in type-II models) Yukawa couplings. The sign
of the correction depends on the sign of $\hat{\lambda}$, which is
in general unknown but is determined in supersymmetric models or
other models where the physics constraints determine the sign (see
below).

In Ref. \cite{wise}, parameters such as $g_1 \sim 2$ (note that we
have changed the sign of $g_1$ to reflect the sign correction in
the Yukawa and the $S$ vev) and $\eta_d \sim 20$ were considered,
corresponding to large $\tan \beta$ and moderate $\hat{\lambda}$.
For these parameters the total width could change substantially,
being corrected by a factor of 121 for Higgs mass of order a TeV.
If, on the other hand parameters were $g_1=-1$ and $\eta_D=-10$,
the branching ratio was down by 0.008. \cite{wise} imagined that
the bottom coupling was changed and the  $\eta_d$-enhanced
deviation from the Standard Model may or may not apply to the
$\tau$ as in Type II models.

Note that for either sign of the correction,  the rate of decay of
the light Higgs into $b$ quarks and hence the total width and the
branching fractions into other modes (in the light Higgs regime
where decays to $b$s dominate) will deviate from the Standard
Model predictions. In Type II models where leptons and down type
quarks both couple to $H_1$,  the best measurement of this Yukawa
deviation at the LHC will be the relative branching ratios of
photons and taus. In other models in which the tau Yukawa is not
changed directly by a large amount  but only through the change in
total width (as might happen for more general MFV models or in
Type III models where the up-type Higgs couples to leptons),  one
would need to measure the absolute decay rate into $\tau$s or
photons since both branching fractions change indirectly through
the change in the Higgs total width.   In this case,  the tau rate
would increase or decrease when the photon rate does, unlike Type
II models where they would change in opposite directions.  The
ratio of photon and tau partial branching fractions   is likely to
be measured at the 15-30 \% level \cite{zeppenfeld, chgeneral} and
absolute branching fractions might also be measured at reasonable
levels \cite{zeppenfeld}. Of course especially for the photon loop
effects from nonstandard model physics might also be significant.
In addition, radiative effects involving $b$s might further
suppress this decay \cite{suppressedb} as we further discuss
below. Whether or not radiative effects are significant, tree
level effects can dominate and give rise to deviations from the
expected Standard Model ratios at a potentially measurable level
for the LHC and a readily attainable level for the ILC.

Notice that the  results are very similar to those from
\cite{wise} since the (large) corrections to the down type Yukawa
coupling match. The difference would be only in the up type
Yukawas, where the \cite{wise} Lagrangian has a correction
$\eta_u$ which is in principle independent of $\eta_d$. However,
since this is small by assumption, it  won't make any measurable
difference.

It is also useful to note what happens to Yukawa modifications
when a $Z_2$ symmetry is preserved by the quartic interactions
that would forbid $\lambda_6$ and $\lambda_7$. In that case, the
$\tan\beta$ enhancement no longer exists, since $\hat{\lambda}$ is
proportional to $\cos\beta$. This is in fact what happens in the
MSSM. Although this can decouple more quickly than without
$\tan\beta$ enhancement as has been noted in several places (see
\cite{gunionhaber,howie} for example), and is not an enhancement
that would allow the sort of large change in branching ratio that
was considered in Ref. \cite{wise}, it still might be measurable.

For example, from Eq. (\ref{lambdahat}), we can deduce the
tree-level change in Yukawa in a supersymmetric theory, which is
${g^2+g'^2 \over 2} {v^2 \over m_H^2}$, which is about $0.3$ for
Higgs mass comparable to $v$. The LHC will measure couplings, even
for the tau, to an accuracy of at most about 15 \%
\cite{carena,ilc,zeppenfeld}. This means that a 2 sigma
measurement might just probe this deviation from the Standard
Model. Our calculation would have to be performed more reliably in
the limit that the second Higgs is light enough to generate a
measurable deviation in Yukawas, but is probably reasonably
accurate since the expansion really involves the light Higgs mass
squared divided by the heavy Higgs mass squared. We leave more
detailed study with light second Higgs mass for future work.

At the ILC, both $b$ and $\tau$ partial widths will be well
measured, with the $b$ partial width particularly accurate. The
anticipated experimental accuracy in the $b$ width will be between
1 and 2.4\%, that for the $\tau$ is between 4.6 and 7.1\%, for the
photons is between 23 and 35\%, and for the $c$ is 8.1-12.3\% (see
Ref. \cite{ilcreport} and references therein). These numbers do
not include the theoretical uncertainties estimated to be about
2\% for the $b$s and 12\% for the $c$s for example
\cite{chgeneral}.  Note that the best measured mode at the ILC,
the $b$ decay mode, is most likely to  have a Yukawa that deviates
from its Standard Model prediction. A sufficiently accurate
measurement of the total width will also probe deviations of the
decay width to $b$s when that mode dominates. Clearly, by
measuring these relative rates at the ILC one can hope to explore
much higher masses indirectly through precision light Higgs
studies. This could be a very interesting probe of higher-energy
physics than will be directly accessible.

Notice also that the radiative corrections for very large
$\tan\beta$ in supersymmetric theories can take the opposite sign
to the tree level corrections we have considered here, as analyzed
in \cite{suppressedb}. If $\tan\beta$ is indeed very large these
radiative corrections need to be taken into account and can end up
suppressing the $b$ branching fraction.

It is straightforward to extend our analysis to the lepton sector.
We consider models that preserve a discrete $Z_2$ flavor symmetry
so that only one type of Higgs field has a tree-level coupling to
each of the different fermion types.
Clearly, only in Model II, where we expect the leptons to couple
to $H_1$ as do the down quarks,  do we expect $\tan\beta$
enhancement in the lepton Yukawas. In these models
we would expect the $\tau$ branching fraction and $b$ branching
fraction to change a comparable amount (up to loop effects).
Radiative corrections to the bottom can be much bigger than those
to the tau \cite{babu} (see also \cite{largetanbeta}, but unless
$\tan\beta$ is very large these are generally smaller than the
tree-level corrections but eventually should be accounted for as
well.

In Model I, where only a single Higgs participates in the Yukawas,
we expect $H_2$ to couple to all fermions or else the top quark
mass would be too low, which means that no fermions would get
large Yukawa corrections. In Type III models as well, the leptons
couple to $H_2$. In both of these latter cases, the correction is
suppressed by a factor of $\cot\beta$ and will be too small to
matter in the large $\tan\beta$ limit.

\section{Gauge Boson Coupling}

It is also interesting to consider the light Higgs to two $W$
coupling since the growth with energy isn't fully stopped until we
reach the second Higgs. This is similar to the analysis of
\cite{rattazzi} where it was argued there would be growth with
energy in $W W$ scattering until the strongly interacting scale in
composite Higgs models. In practice at the LHC this will probably
be a less promising way to search for evidence of a second Higgs
  because the $H \to WW$ won't be sufficiently
precisely measured since the energy reach isn't big enough to
enhance the cross section sufficiently,  and because in the case
of a doublet Higgs field the corrections to the scattering
effectively arise from higher-dimension operators than in Ref.
\cite{rattazzi}.

One way to understand the source of the correction to the Higgs
$WW$ coupling  in the strong coupling case \cite{rattazzi} is from
a higher order operator of the form
\begin{equation}
{ c_H \over 2 f^2}  {v^2 \over f^2} \partial_\mu (H^\dagger H)
\partial^\mu (H^\dagger H)
\end{equation}
where $f$ is the scale of strong physics,  which, after a shift in
$H$ field gives a correction
\begin{equation}
c_H m_W^2 {h \over v} W_\mu W^\mu
\end{equation}
where $h$ is the light Higgs.  In effect, a dimension-6 operator
could arise only in the presence of a singlet or triplet to be
exchanged. In our case, with only a doublet Higgs, our correction
is higher order. We expect a correction to $h WW$ of order
$(v/m_H)^4$.

In practice, we know precisely the coupling of $h$ to a pair of
$W$s. It is proportional to
$\sin(\alpha-\beta)=1-{\cos^2(\alpha-\beta) \over 2}$. The
correction to unityh is indeed suppressed by  $(v/m_H)^4$ as we
expected and is likely to be too small to measure.

\section{Heavy Higgs Direct Searches}

The heavy Higgs two vector boson coupling is suppressed by
$\cos(\alpha-\beta)$, since the vev of the field is suppressed by
this factor and the vev enters the single Higgs two gauge boson
coupling. This means that even when the two gauge boson decay is
kinematically allowed, it won't generally dominate. Similarly,
heavy Higgs boson production  is suppressed.\footnote{Here we are
neglecting the other Higgs states but these will also be difficult
to find.} Notice in the coupling to two $W$s there are no
compensating $\tan\beta$ factors as there were for the down and
potentially $\tau$ Yukawa corrections so the heavy Higgs to two
gauge boson coupling is indeed small.

CMS recently (2007) \cite{cmstdr} studied the heavy Higgs
discovery reach in the MSSM with systematic uncertainties taken
into account. They found for a relatively light second Higgs (CP
even or odd) that to find a Higgs of 150 GeV, $\tan\beta$ must be
greater than about 16 and for  a Higgs of 250 GeV, $\tan \beta$
must be greater than about 35. This can be compared to the results
from the Atlas TDR from 1999 \cite{atlastdr} quoted by
\cite{carena} where it was already noted that for Higgs mass of
250 GeV, $\tan\beta$ greater than 8 was necessary whereas for 500
GeV $\tan \beta$ needs to be at least 17. Clearly the situation
has become worse with better understanding of the systematics and
a reasonably large value of $\tan\beta$ is required to discover
the heavy Higgs.

The required large value of $\tan\beta$ is readily understood from
our earlier considerations. In Type II two Higgs models preserving
a $Z_2$ symmetry, large $\tan\beta$ tells us the coupling of the
heavy Higgs to bottom type quarks is enhanced whereas the coupling
to top quarks and two gauge bosons is suppressed. Therefore
production through bottoms is enhanced when $\tan\beta$ is large.
Moreover decays to taus are enhanced in this limit as well and
that is likely to be the best search mode. Note that even with the
$\tan \beta$-enhanced coupling to bottom quarks, the amplitude is
proportional to  the bottom Yukawa as well
  so only when $\tan \beta $ is sufficiently sizable
will the production and decay become visible.

Notice that although the analysis was done for the MSSM, the
answer can be readily taken over to more general two Higgs models.
The bottom and top Yukawas will be determined by $\tan\beta$ at
leading order. The more model-dependent coupling
 is the coupling to two
gauge bosons which is suppressed by the heavy Higgs vev (or
equivalently $\cos(\beta-\alpha)$). Once this is sufficiently
small neither production nor decay through this mode is relevant.

Note that Ref. \cite{wise} considered particular parameters in the
two Higgs model to show that a heavy Higgs (of order TeV) can
readily elude detection but induce large deviations from the
Standard Model in the low-energy effective theory. They had large
bottom Yukawa and small top Yukawa (to suppress standard Higgs
production channel). Our point is that this happens automatically
for large $\tan\beta$ (but not so large that the Higgs will be
produced directly). Furthermore the CMS analysis shows that even a
much lighter heavy Higgs  than considered in \cite{wise} will not
be seen unless $\tan\beta$ is sufficiently large. Of course even
if $\tan\beta$ is large and the second Higgs is discovered, it
will still  be worthwhile to explore the types of deviations in
Yukawas we have considered.

We conclude that there is a large region of parameter space where
precision light Higgs decays will be the best way to search for
evidence of a second Higgs.  This can also be a way of
distinguishing among higher-dimension operator contributions to
the Higgs mass squared as we discuss in the following section.

\section{Implications for Testing Higher Dimension Operators}

 Recently Ref. \cite{seiberg}  suggested the
existence of higher dimension operators involving Higgs fields as
a way of summarizing all possible models that might raise the
Higgs mass without a large stop (or $A$ term) (see also
\cite{jose,strumia}) in models that didn't contain new light
fields into which the Higgs could decay and escape observation
(see \cite{gunion, neal,kaplan,Accomando:2006ga} and references
therein). In this way they hoped to address the little hierarchy
problem that seems to require a large stop mass to raise the Higgs
mass adequately. It is of interest to ask how to detect such
higher dimension operators.

The obvious hope would be to find and measure additional Higgs
states and study the mass relations. However, as we have
discussed, it will be difficult to find a second Higgs over much
of the relevant parameter range and similar considerations apply
to other states from the Higgs sector. This leaves the question if
the light Higgs does indeed have bigger mass than expected on the
basis of the MSSM, are there other ways to distinguish among
different possible higher dimension operators that might be
contributing to its mass? Here we show that the likely leading
operator to affect the Higgs mass is also precisely the one that
should be best tested in the Higgs partial widths and the  Yukawa
analysis above readily applies. That means that not only can
studying the branching fractions test for these operators, it
could help distinguish among them.

In Ref. \cite{seiberg}, it was demonstrated that at leading order
in an effective dimension expansion, only one operator contributes
to the light Higgs mass in the large $\tan\beta$  (but not so
large that higher order mass suppressed terms dominate over
$\cot\beta$ suppressed terms) limit. This operator is

\begin{equation}
{\lambda \over M} \int d^2 \theta (H_u H_d)^2
\end{equation}

(For simplicity, we assume all new parameters are real. We are
also retaining the notation of Ref. \cite{seiberg} where $H_u$ and
$H_d$ are  used for $H_2$ and $H_1$ respectively.) When combined
with the supersymmetric operator $\int d^2 \theta \mu H_u H_d$, we
find the quartic term

\begin{equation}
{2 \lambda \mu \over M} (H_u H_d) (H_u^\dagger H_u + H_d^\dagger
H_d)
\end{equation}
Such a term can also arise from a $D$-term type interaction.

Defining  $\epsilon_1=\lambda \mu/M$, one finds a Higgs mass
correction of order $8 \epsilon_1 \cot\beta v^2$ (with the $v=246
{\rm GeV}$ convention we have been using) \cite{seiberg}
 The authors of Ref.
\cite{seiberg} argued that one can get a sufficiently large
correction to the Higgs mass (one that replaces the large stop
contribution) for parameters such as $\tan\beta \sim 10$ and
$\epsilon_1\sim 0.06$

Notice the interesting feature of this operator. Even though the
only breaking of the $Z_2$ symmetry in the superpotential was
through the lower-dimension $\mu$-term, it feeds into a
dimension-4 $Z_2$-violating operator in the potential. This
$Z_2$-breaking, characterized by $\mu/M$, can be sizable.  This
will be important below.

Alternatively, $\tan\beta$ could be so big  (hence $\cos\beta$ so
small) that terms suppressed by more powers of $1/M$ dominate over
the leading $1/M$ correction.  Such an unsuppressed contribution
might arise from
 an operator $(H_u^\dagger H_u)^2$ for example.

We can now use our previous analysis to consider the effect of
such operators on a light Higgs coupling to down-type quarks and
charged leptons. We see that the first operator, while suppressed
by $\cot\beta$ in its impact on the Higgs mass, is in fact exactly
the type of operator that gets a $\tan \beta$ {\it enhanced}
contribution to the  Yukawa coupling deviations above. That is
because it arises through the $Z_2$-violating $\mu$-term and
contributes directly to $\lambda_7$. In particular, $\lambda_7\sim
2 \epsilon_1$. If $\tan \beta $ is large, Yukawa couplings receive
corrections from $\tan\beta 2 \epsilon_1 (v/m_H)^2$ effects. As an
example, if
 $m_H\sim 1.5 v$, with the parameters given above,
 these contributions could  reduce the $h\to\tau \bar{\tau}$
rate by a factor of 4, while  increasing the $h\to\gamma \gamma $
rate by a similar factor (due to the decreased rate to $b
\bar{b}$). Even without discovering the second Higgs, these
effects could be big enough to test for the higher dimension
operators indirectly.

As an aside we note that recent papers \cite{gunion, neal,kaplan}
have considered the possibility that the light Higgs does not
decay into the modes that have been sought for at LEP. In those
papers there were alternative beyond the minimal supersymmetric
model light modes available into which the Higgs can decay. We
have just seen that even without these additional light modes, the
Higgs branching ratio into    $b \bar{b}$ and $\tau \bar{\tau}$
can be reduced substantially. However even when the branching
ratio to $b$s is so reduced that other modes dominate, the
alternative decay modes would have been visible as well, so the
Higgs mass bound would not be reduced by more than a few GeV
\cite{lep} so this doesn't alter the allowed range of $M$
significantly.

Returning to the effects on LHC branching fractions, for an
operator whose contribution to the squared Higgs mass is
suppressed by two powers of $M$ but not by $\cos\beta$ such as
$(H_u^\dagger H_u)^2$, the contribution to the deviation in the
Yukawa will nonetheless be suppressed by $\cos\beta$. Therefore
the effect on the bottom Yukawa is much smaller than for the
$Z_2$-breaking operator we just considered.

We can readily understand the relative signs and magnitudes of
Yukawa corrections from the various operators by studying the sign
and $\cos\beta$ dependence contributions to both the light Higgs
mass squared and to the bottom-type Yukawa couplings of the
various operators in the limit that $\cos\beta$ is small.
\\

\begin{tabular}{|r|r|r|}
\hline
Operator&Mass Squared Contribution&Down Yukawa Contribution\\
\hline
$(H_u H_d)^2$&$\cos^2\beta$&$\cos\beta$\\
$ (H_u^\dagger H_u)(H_u H_d)$&$\cos\beta$&1\\
$(H_u^\dagger H_u)^2$&1&$-\cos\beta$\\
\hline
\end{tabular}
\\
\\
We see that the operators consistent with the $Z_2$ symmetry do
indeed give $\cos\beta$-suppressed contributions to the change in
the down-type Yukawas.  We also see that the effect of the last
operator has the opposite sign which is why it increases the
branching fraction of the bottom whereas the other operators
decrease it. This should be a powerful tool for distinguishing
among higher dimension operators should they be present.

Therefore if a light Higgs consistent with current experimental
constraints and small stop mass is discovered (assuming small
$A$), measuring branching ratios could test which higher dimension
operator is the relevant one in raising its mass. In particular
the effects of the first type of operator can have significant
effects on the Higgs decay rate and branching ratios which we
would not expect for the higher effective dimension operators.

\section{Conclusion}

We conclude that is is very likely that even if there are two
Higgs doublet fields and the second neutral Higgs scalar is
kinematically accessible to the LHC, it is likely that the second
Higgs will elude direct detection. This makes the question of
indirect evidence for the full Higgs sector very important.

We have seen that there is a large parameter range where precision
measurements, in particular of the branching fraction of the light
Higgs into taus vs. photons, can find indirect evidence for a
second Higgs field. If there are $Z_2$-violating interactions in
the Higgs quartic terms, there can be enormous changes to the
bottom and tau branching fractions, so large that they will be
reflected in the overall Higgs decay rate and will result in a
significant change in the  branching fraction to other modes.

We have also seen that such measurements can  be a powerful way to
test for higher dimension operators  in a supersymmetric theory
and that the operator which is perhaps most likely to affect the
light Higgs mass will yield significant changes to the decay
widths into bottoms and taus.

Therefore precision Higgs branching fraction measurements can be
extremely important if the world does in fact contain two Higgs
fields. It will be interesting to do more detailed explorations of
parameters, to consider which range is most natural and for how
large a parameter range the considerations of this paper apply. It
will also be of interest to incorporate the effects of CP
violation.

\section{Acknowledgements} I thank Christophe Grojean and Cedric Delaunay for
participating in the early stages of this work and I thank them
and Liam Fitzpatrick, Spencer Chang, Howie Haber, Gilad Perez,
Massimo Porrati, Giovanni Villadoro, and Mark Wise for reading the
manuscript and for very useful discussions.   This work was
supported in part by NSF grants PHY-0201124 and PHY-0556111.
 I also thank
NYU for their hospitality while this work was being completed.

\newpage


\begin{thebibliography}{9}

\bibitem{rattazzi}
  G.~F.~Giudice, C.~Grojean, A.~Pomarol and R.~Rattazzi,
  JHEP {\bf 0706}, 045 (2007)
  [arXiv:hep-ph/0703164].



\bibitem{wise}
  S.~Mantry, M.~Trott and M.~B.~Wise,
  arXiv:0709.1505 [hep-ph].

  \bibitem{gunionhaber}
  J.F.~Gunion and H.E.~Haber,
Phys.\ Rev. {\bf D67} (2003) 075019 [arXiv:hep-ph/0207010].



  \bibitem{hhguide}
  J.F.~Gunion, H.E.~Haber, G.L.~Kane, and S.~Dawson,
{\it The Higgs Hunter's Guide} (Perseus Publishing, Cambridge, MA,
1990)

  \bibitem{mfv}
  R.~S.~Chivukula and H.~Georgi,
  Phys.\ Lett.\  B {\bf 188}, 99 (1987).
  L.~J.~Hall and L.~Randall,
  Phys.\ Rev.\ Lett.\  {\bf 65}, 2939 (1990).
  G.~D'Ambrosio, G.~F.~Giudice, G.~Isidori and A.~Strumia,
  Nucl.\ Phys.\  B {\bf 645}, 155 (2002)
  [arXiv:hep-ph/0207036].
  V.~Cirigliano, B.~Grinstein, G.~Isidori and M.~B.~Wise,
  Nucl.\ Phys.\  B {\bf 728}, 121 (2005)
  [arXiv:hep-ph/0507001].
  A.~J.~Buras,
  Acta Phys.\ Polon.\  B {\bf 34}, 5615 (2003)
  [arXiv:hep-ph/0310208].


  \bibitem{chgeneral}
  M.~Carena and H.~E.~Haber,
  Prog.\ Part.\ Nucl.\ Phys.\  {\bf 50}, 63 (2003)
  [arXiv:hep-ph/0208209].

  \bibitem{howie}
  H.~E.~Haber, M.~J.~Herrero, H.~E.~Logan, S.~Penaranda, S.~Rigolin and D.~Temes,
in {\it Proc. of the 5th International Symposium on Radiative
Corrections (RADCOR 2000) } ed. Howard E. Haber,
  arXiv:hep-ph/0102169,
  H.E. Haber, M.J. Herrero, H.E. Logan,
S. Pe\~naranda, S. Rigolin and D. Temes, {Phys. Rev.} {\bf D63},
055004 (2001) [arXiv:hep-ph/0007006].

  \bibitem{pandita}
  P.~N.~Pandita,
  Phys.\ Lett.\  B {\bf 151}, 51 (1985).

  \bibitem{davidson}
   S.~Davidson and H.~E.~Haber,
  Phys.\ Rev.\  D {\bf 72}, 035004 (2005)
  [Erratum-ibid.\  D {\bf 72}, 099902 (2005)]
  [arXiv:hep-ph/0504050],
  [Erratum: {\bf D72}, 099902 (2005)],
H.E.~Haber and D.~O'Neil, {Phys.\ Rev.} {\bf D74}, 015018 (2006)
[Erratum:
 {\bf D74}, 059905 (2006)] [arXiv:hep-ph/0602242]

  \bibitem{carena}
  M.~Carena, H.~E.~Haber, H.~E.~Logan and S.~Mrenna,
  Phys.\ Rev.\  D {\bf 65}, 055005 (2002)
  [Erratum-ibid.\  D {\bf 65}, 099902 (2002)]
  [arXiv:hep-ph/0106116].

  \bibitem{ilc}
  S.~Heinemeyer {\it et al.},
  arXiv:hep-ph/0511332.

  \bibitem{babu}
  K.~S.~Babu and C.~F.~Kolda,
  Phys.\ Lett.\  B {\bf 451}, 77 (1999)
  [arXiv:hep-ph/9811308].

  \bibitem{largetanbeta}
  D.~M.~Pierce, J.~A.~Bagger, K.~T.~Matchev and R.~j.~Zhang,
  Nucl.\ Phys.\  B {\bf 491}, 3 (1997)
  [arXiv:hep-ph/9606211].
  L.~J.~Hall, R.~Rattazzi and U.~Sarid,
  Phys.\ Rev.\  D {\bf 50}, 7048 (1994)
  [arXiv:hep-ph/9306309].
  M.~Carena, M.~Olechowski, S.~Pokorski and C.~E.~M.~Wagner,
  Nucl.\ Phys.\  B {\bf 426}, 269 (1994)
  [arXiv:hep-ph/9402253].










  \bibitem{zeppenfeld}
  M.~Duhrssen, S.~Heinemeyer, H.~Logan, D.~Rainwater, G.~Weiglein and D.~Zeppenfeld,
  arXiv:hep-ph/0407190.

  \bibitem{suppressedb}
   M.~Carena, S.~Heinemeyer, C.~E.~M.~Wagner and G.~Weiglein,
  ``Suggestions for improved benchmark scenarios for Higgs-boson searches  at
  LEP2,''
  arXiv:hep-ph/9912223.
  CITATION = HEP-PH/9912223;
  M.~Carena, S.~Mrenna and C.~E.~M.~Wagner,
  Phys.\ Rev.\  D {\bf 60}, 075010 (1999)
  [arXiv:hep-ph/9808312].
  W.~Loinaz and J.~D.~Wells,
  Phys.\ Lett.\  B {\bf 445}, 178 (1998)
  [arXiv:hep-ph/9808287].
  M.~Carena, J.~R.~Ellis, A.~Pilaftsis and C.~E.~M.~Wagner,
  Nucl.\ Phys.\  B {\bf 586}, 92 (2000)
  [arXiv:hep-ph/0003180],
  K.~S.~Babu and C.~F.~Kolda,
  Phys.\ Lett.\  B {\bf 451}, 77 (1999)
  [arXiv:hep-ph/9811308].
  \bibitem{Balazs:1998nt}
  C.~Balazs, J.~L.~Diaz-Cruz, H.~J.~He, T.~Tait and C.~P.~Yuan,
  Phys.\ Rev.\  D {\bf 59}, 055016 (1999)
  [arXiv:hep-ph/9807349].




  \bibitem{cmstdr}
  A.~Ball, M.~Della Negra, A.~Petrilli and L.~Foa  [CMS Collaboration],
  J.\ Phys.\ G {\bf 34} (2007) 995.

  \bibitem{atlastdr}
  ``ATLAS: Detector and physics performance technical design report. Volume
  1 (1999),''
  CITATION = ATLAS-TDR-14;

  \bibitem{ilcreport}
  A.~Djouadi, J.~Lykken, K.~Monig, Y.~Okada, M.~J.~Oreglia and S.~Yamashita,
  arXiv:0709.1893 [hep-ph].




\bibitem{seiberg}
  M.~Dine, N.~Seiberg and S.~Thomas,
  arXiv:0707.0005 [hep-ph].

  \bibitem{jose}
  A.~Brignole, J.~A.~Casas, J.~R.~Espinosa and I.~Navarro,
  Nucl.\ Phys.\  B {\bf 666}, 105 (2003)
  [arXiv:hep-ph/0301121].

  \bibitem{strumia}
  A.~Strumia,
  Phys.\ Lett.\  B {\bf 466}, 107 (1999)
  [arXiv:hep-ph/9906266].


  \bibitem{gunion}
  R.~Dermisek and J.~F.~Gunion,
  Phys.\ Rev.\ Lett.\  {\bf 95}, 041801 (2005)
  [arXiv:hep-ph/0502105].



  \bibitem{neal}
  S.~Chang and N.~Weiner,
  arXiv:0710.4591 [hep-ph].

  \bibitem{kaplan}
  L.~M.~Carpenter, D.~E.~Kaplan and E.~J.~Rhee,
  arXiv:hep-ph/0607204.

  \bibitem{Accomando:2006ga}
  E.~Accomando {\it et al.},
  arXiv:hep-ph/0608079.


  \bibitem{lep}
  P.~Achard {\it et al.}  [L3 Collaboration],
  Phys.\ Lett.\  B {\bf 583}, 14 (2004)
  [arXiv:hep-ex/0402003].


\end{thebibliography}
\end{document}